# The Yeast Cell-Cycle Network Is Robustly Designed

Fangting Li [*†], Tao Long [*†], Ying Lu [*†], Qi Ouyang [*‡] and Chao Tang [*‡§]

[*]Centre for Theoretical Biology and Department of Physics, Peking University, Beijing 100871, China; and [§]NEC Laboratories America, 4 Independence Way, Princeton, NJ 08540, USA

[†]F.L., T.L. and Y.L. contributed equally to this work.

[‡]To whom correspondence should be addressed. E-mail: qi@pku.edu.cn or tang@nec-labs.com.

**Abstract**

**The interactions between proteins, DNA, and RNA in living cells constitute molecular networks that govern various cellular functions. To investigate the global dynamical properties and stabilities of such networks, we studied the cell-cycle regulatory network of the budding yeast. With the use of a simple dynamical model, it was demonstrated that the cell-cycle network is extremely stable and robust for its function. The biological stationary state—the $G_1$ state—is a global attractor of the dynamics. The biological pathway—the cell-cycle sequence of protein states—is a globally attracting trajectory of the dynamics. These properties are largely preserved with respect to small perturbations to the network. These results suggest that cellular regulatory networks are robustly designed for their functions.**

Despite the complex environment in and outside of the cell, various cellular functions are carried out reliably by the underlying biomolecular networks. How is the stability of a cell state achieved? How can a biological pathway take the cell from one state to another reliably? Evolution must have played a crucial role in the selection of the architectures of these networks for them to have such a remarkable property. Much attention has recently been focused on the "topological" properties of large-scale networks (1-5). It was argued that a power-law distribution of connectivity, which is apparent for some bionetworks (2,4), is more tolerable against random failure (1). Here we address this question from a dynamic systems point of view. We study the network regulating the cell cycle of the budding yeast, investigating its global dynamical property and stability. We find that the stationary states of the cell, or states at the checkpoints in general, correspond to global attractors of the dynamics—almost all initial protein states flow to these biological stationary states. Furthermore, the biological pathway of the cell-cycle sequence—which is a particular trajectory in the state space—is a globally stable and attracting trajectory of the dynamics. These dynamic properties, arising from the underlying network connection, are also robust against small perturbations to the network. They are directly responsible for the robustness of the cellular process.

**The Yeast Cell-Cycle Network**

The cell-cycle process, by which one cell grows and divides into two daughter cells, is a vital biological process the regulation of which is highly conserved among the eukaryotes (6). The process consists of four phases: $G_1$ (in which the cell grows and, under appropriate conditions, commits to division), S (in which the DNA is synthesized and chromosomes replicated), $G_2$ (a "gap" between S and M), and M (in which chromosomes are separated and the cell is divided into two). After the M phase, the cell enters the $G_1$ phase, hence completing a "cycle". The process has been studied in great details in the budding yeast *Saccharomyces cerevisiae*, a single-cell model eukaryotic organism (see the Supporting Information for references). There are about 800 genes involved in the cell-cycle process of the budding yeast (7). However, the number of key regulators that are responsible for the control and regulation of this complex process is much smaller. Based on extensive literature studies, we have constructed a network of key regulators that are known so far, as shown in Fig. 1A (details in the Supporting Information).

There are four classes of members in this regulatory network: cyclins (Cln1,2,3 and Clb1,2,5,6, which bind to the kinase Cdc28); the inhibitors, degraders, and competitors of the cyclin/Cdc28 complexes (Sic1, Cdh1, Cdc20, Cdc14); transcription factors (SBF, MBF, Mcm1/SFF, Swi5); and checkpoints (the cell size, the DNA replication and damage, and the spindle assembly). Green arrows represent positive regulations. For example, under rich nutrient conditions and when the cell grows large enough, the Cln3/Cdc28 will be "activated", which in turn activates (by phosphorylation) a pair of transcription factor groups SBF and MBF, which transcriptionally activate the genes of the cyclins Cln1,2 and Clb5,6, respectively. Red arrows represent "deactivation" (inhibition, repression, or degradation). For example, the protein Sic1 can bind to the Clb/Cdc28 complex to inhibit its function, Clb1,2 phosphorylates Swi5 to prevent its entry into the nucleus, while Cdh1 targets Clb1,2 for degradation. The cell-cycle sequence starts when the cell commits to division by activating Cln3 (the START). The subsequent activity of Clb5,6 drives the cell into the S phase. The entry into and exit from the M phase is controlled by the activation and degradation of Clb1,2. After the M phase, the cell comes back to the stationary $G_1$ phase, waiting for the signal for another round of division. Thus the cell-cycle process starts with the "excitation" from the stationary $G_1$ state by the "cell-size" signal and evolves back to the stationary $G_1$ state through a well-defined sequence of states.

**The Model and Dynamic Properties**

In principle, the arrows in the network have very different time scales of action, and a dynamic model would involve various binding constants and rates (8,9). However, since in the cell-cycle network much of the biology seems to be reflected in the on-off characteristics of the network components and we are mainly concerned here with the overall dynamic properties and the stability of the network, we use a simplified dynamics on the network, which treats the nodes and arrows as logic-like operations[¶]. Thus, in the model each node $i$ has only two states, $S_i=1$ and $S_i=0$, representing the active and the inactive state of the protein, respectively. The protein states in the next time step are determined by the protein states in the present time step via the following rule:

$$S_i(t+1) = \begin{cases} 1, & \sum_j a_{ij} S_j(t) > 0 \\ 0, & \sum_j a_{ij} S_j(t) < 0 \\ S_i(t), & \sum_j a_{ij} S_j(t) = 0 \end{cases} \quad (1)$$

where $a_{ij} = a_g$ for a green arrow from protein $j$ to protein $i$ and $a_{ij} = a_r$ for a red arrow from $j$ to $i$. We first focus on the case where all the other checkpoints are always off, except that of the cell size. That is, the cell size checkpoint will act as a start signal, while other checkpoints will always let the "traffic" pass when it come to it. We therefore arrive at a slightly simplified network shown in Fig. 1B, with 11 nodes (plus a signal node). We have also added "self-degradation" (yellow loops) to those nodes that are not negatively regulated by others[∥]. The degradation is modeled as a time-delayed interaction: if a protein with a self yellow arrow is active at time $t$ ($S_i(t)=1$) *and* if its total input is zero from $t+1$ to $t=t+t_d$, it will be degraded at $t=t+t_d$, i.e. $S_i(t+t_d)=0$. The results presented below were obtained with $a_g=-a_r=1$ and $t_d=1$. As will be discussed later, the overall dynamic properties of the network are not very sensitive to the choice of these parameters.

**Fixed Points.** Using the dynamic model described above, we study the time evolution of the protein states. First, we study the attractors of the network dynamics by starting from each of the $2^{11}$=2048 initial states in the 11-node network of Fig. 1B. We find that all of the initial states eventually flow into one of the 7 stationary states (fixed points), shown in Table 1. Among the 7 fixed points, there is one big fixed point attracting 1764 or about 86% protein states. Remarkably, this super stable state is the biological $G_1$ stationary state. The advantage for a cell's stationary state to be a big attractor of the network is obvious—the stability of the cell state is guaranteed. Under normal conditions, the cell will be sitting at this fixed point, waiting for the signal for another round of division.

**Biological Pathway.** Next, we start the cell-cycle process by "exciting" the $G_1$ stationary state with the cell size signal, and observe that the system goes back to the $G_1$ stationary state. The temporal evolution of the protein states, presented in Table 2, indeed follows the cell-cycle sequence, going from the excited $G_1$ state (the START) to the S phase, the $G_2$ phase, the M phase, and finally to the stationary $G_1$ state. This is the biological trajectory or pathway of the cell-cycle network.

To investigate the dynamical stability of this biological pathway, we study the dynamic trajectories of all 1764 protein states that will flow to the $G_1$ fixed point. In Fig. 2, each of these protein states is represented by a dot, with the arrows between them indicating dynamic flows from one state to another. The biological pathway is colored in blue and so is the node representing the $G_1$ stationary state. We see that the dynamic flow of the protein states is convergent onto the biological pathway, making the pathway an attracting trajectory of the dynamics. With such a topological structure of the phase diagram of protein states, the cell-cycle pathway is a very stable trajectory—it is very unlikely for a sequence of events, starting at the beginning (or at any other point) of the cell-cycle process, to deviate from the cell-cycle pathway. Interestingly, the topology of the converging trajectories shown in Fig. 2 is reminiscent of the converging kinetic pathways in protein folding where a protein sequence is facing the challenge of finding the unique native state among a huge number of conformations (10-12).

**Comparison with Random Networks.** To investigate how likely a big fixed point and a converging pathway can arise by chance, we study an ensemble of random networks (13,14) that have the same numbers of nodes and links in each color as in the cell-cycle network. We find that random networks typically have more attractors (fixed points and limit cycles), with the average number being 14.28. The sizes of the basins of attraction in the random networks have a power-law distribution, as shown in Fig. 3A. The probability for a random network to have an attractor of a basin size $B$ equal to or larger than that of the cell-cycle network ($B \geq 1764$) is 10.34%.

To quantify the "convergence" of trajectories, we define a quantity $w_n$ ($n$=1, 2, …, 2048) for each of the 2048 network states that measures the overlap of its trajectory with all other trajectories (Fig. 3C). Denote $T_{j,k}$ the total traffic flow through the arrow $A_{j,k}$ that takes state $j$ to $k$ in one time step, *i.e.* $T_{j,k}$ is the total number of trajectories starting from all network states that pass through $A_{j,k}$. If the trajectory from $n$ to its attractor has $L_n$ steps so that it consists $L_n$ arrows $A_{k-1,k}$, $k=1,2, \ldots, L_n$, $w_n = \sum_{k=1}^{L_n} T_{k-1,k} / L_n$. The overall overlap of all trajectories in a network can be measured by $W=\langle w_n \rangle$, where the average is over all network states. The normalized histogram of $w_n$ for all network states is shown in Fig. 3B, for both the cell-cycle and the random networks. Without any significant overlap or convergence of trajectories and with a much shorter transient times to attractors, the random networks have their $w$-distribution peaked at small $w$'s, with an average $W=124$. While for the cell-cycle network, the distribution is peaked at very large numbers ($W=743$), indicating significant convergence of trajectories. The probability for a random network to have a $W \geq 743$ is 0.25%.

**Network Perturbations.** We see that the cell-cycle network has two distinct dynamic properties compared with random networks: it has a super fixed point and it has a converging pathway. What effects would perturbations of the network have on these properties? We perturbed a network by deleting an interaction arrow, adding a green or red arrow between nodes that are not linked by an arrow, or switch a green arrow to red and vice versa. The relative change in the size $B$ of the basin of attraction for the biggest attractor, $\Delta B/B$ were then measured as a result of the perturbation. The distributions of $\Delta B/B$ are plotted in Fig. 4 for each kind of the perturbations, respectively, along with those obtained from the ensemble of random networks. We observe that only a very small fraction of perturbations will eliminate the fixed point completely ($\Delta B/B=1$). For most perturbations, the relative changes of the basin size are small. A similar behavior in the changes of the quantity $W$ as results of the perturbations was also seen. Interestingly, this high "homeostatic stability" (13) is also evident in the ensemble of random networks of the same size (Fig. 4). In fact, we found that for random networks with the dynamic rule of Eq. 1 the homeostatic stability increases monotonically with the average number of arrows per node $k$ (15), which is very different from the random Boolean network where a "chaotic" phase with low homeostatic stability is seen for $k > k_c$ (13). Recent studies suggest that either a scale-free Boolean network (16) or a genetic network with minimal frustration (17) would also lead to a more stable phase.

To examine the effects of these perturbations on the biological pathway itself, for each perturbed cell-cycle network we start at the START state and follow its time evolution. We found that under perturbation a significant fraction of the trajectories reach the $G_1$ stationary state and the cell-cycle sequence is by far the most probable trajectory (Fig. 5).

**Other dynamical rules.** We found that the results are insensitive to the values of the weights $a_g$ and $a_r$ in Eq. 1 and to the protein lifetime $t_d$, as long as $-a_r \geq a_g$ and $t_d > 0$. For example, with $a_r=-10$, $a_g=1$, and $t_d=4$, there exist the same 7 fixed points. The $G_1$ fixed point attracts 90% of all protein states and $W=907$. The network is somewhat more robust against perturbation (see Fig. S-1 in the Supporting Information). Preliminary results with differential equations replacing the simple discrete dynamic rule support the overall conclusions (18).

**Other checkpoints.** We also studied the cases in which one of the other checkpoints, instead of the "cell size", will act as the stop-go signal. We found that in all cases there exists a big fixed point that corresponds to the biological state waiting at the checkpoint and the biopathway is a converging trajectory. The studies were done on the full network (Fig. 1A), keeping only one checkpoint at a time. The basin size $B$ of the big fixed point and the convergence measure $W$ of the biopathway for each checkpoint are, respectively, $B$=99.4% and $W$=4257 (Inter-S), $B$=89.8% and $W$=3821 (Spindle Assembly), $B$=99.8% and $W$=4925 (DNA Damage). For comparison, the corresponding values for the Cell-Size checkpoint with the full network are $B$=90.8% and $W$=6757 (see Fig. S-2 in the Supporting Information).

**Discussion**

We have demonstrated that the yeast cell-cycle network is robustly designed. The biological states at the checkpoints are big attractors and the biopathway is an attracting trajectory. These robust dynamical properties are also seen in the life-cycle network of the budding yeast (19), suggesting that they may be common features of regulatory networks. The cell-cycle network is rather stable against perturbations. Note that the network we studied (Fig. 1A) is only a skeleton of a larger cell-cycle network with many "redundant" components and interactions (e.g. any member of the $G_1$ cyclins can, to a large extend, perform the functions of other members). Thus, we expect the complete network to be even more stable against perturbations.

The idea that aspects of biological systems can be modeled as dynamic systems and biological states can be interpreted as attractors has a long history, with examples in neural networks (20,21), immune systems (22,23), genetic networks (24,13), cell regulatory network (25), and ecosystems (26). Our study on an actual yeast cellular network lends support to this idea. Furthermore, our results suggest that not only biological states correspond to big fixed points but the biological pathways are also robust.

Functional robustness has been found in other biological networks, *e.g.* in the chemotaxis of *E. Coli* (in the response to external stimuli) (27) and in the gene network setting up the segment polarity in insects development (with respect to parameter changes) (28,29). It has also been found at the single molecular level—in the mutational and thermodynamic stability of proteins (30). In some sense, biological systems have to be robust in order to function in complex (and very noisy) environments. More robust could also mean more evolvable and thus more likely to survive—a robust "module" is easier to be modified, adapted, added-on, and combined with others for new functions and new environments (31). Indeed, robustness may provide us with a handle to understand the profound driving force of evolution.

**Acknowledgements**

This work was partly supported by the National Key Basic Research Project of China (No.2003CB715900). T.L. and Y.L. acknowledge the support from the Jun Zheng foundation. We thank Hamid Bolouri, Terry Hwa, Stuart Kauffman, Hao Li, Albert

**Footnotes**

¶Making the time constants of all arrows the same could have disastrous consequences in network dynamics. However, we are saved for this particular network because of its intrinsic sequential nature. We have tested the dynamics with varied time scales of action (phosphorylation and transcriptional activation) for different arrows and obtained similar results.

∥This is a simplification for the actual degradation processes. See the Supporting Information for details.

| Basin size | Cln3 | MBF | SBF | Cln1,2 | Cdh1 | Swi5 | Cdc20 | Clb5,6 | Sic1 | Clb1,2 | Mcm1 |
|---|---|---|---|---|---|---|---|---|---|---|---|
| 1764 | 0 | 0 | 0 | 0 | 1 | 0 | 0 | 0 | 1 | 0 | 0 |
| 151 | 0 | 0 | 1 | 1 | 0 | 0 | 0 | 0 | 0 | 0 | 0 |
| 109 | 0 | 1 | 0 | 0 | 1 | 0 | 0 | 0 | 1 | 0 | 0 |
| 9 | 0 | 0 | 0 | 0 | 0 | 0 | 0 | 0 | 1 | 0 | 0 |
| 7 | 0 | 1 | 0 | 0 | 0 | 0 | 0 | 0 | 1 | 0 | 0 |
| 7 | 0 | 0 | 0 | 0 | 0 | 0 | 0 | 0 | 0 | 0 | 0 |
| 1 | 0 | 0 | 0 | 0 | 1 | 0 | 0 | 0 | 0 | 0 | 0 |

**Table 1.** The fixed points of the cell-cycle network. Each fixed point is represented in a row. The left column is the size of the basin of attraction for the fixed point; the other 11 columns show the protein states of the fixed point. The protein states of the biggest fixed point correspond to that of the $G_1$ stationary state.

| Step \ Protein | Cln3 | MBF | SBF | Cln1,2 | Cdh1 | Swi5 | Cdc20 & Cdc14 | Clb5,6 | Sic1 | Clb1,2 | Mcm1/ SFF | Phase |
|---|---|---|---|---|---|---|---|---|---|---|---|---|
| 1 | 1 | 0 | 0 | 0 | 1 | 0 | 0 | 0 | 1 | 0 | 0 | START |
| 2 | 0 | 1 | 1 | 0 | 1 | 0 | 0 | 0 | 1 | 0 | 0 | $G_1$ |
| 3 | 0 | 1 | 1 | 1 | 1 | 0 | 0 | 0 | 1 | 0 | 0 | $G_1$ |
| 4 | 0 | 1 | 1 | 1 | 0 | 0 | 0 | 0 | 0 | 0 | 0 | $G_1$ |
| 5 | 0 | 1 | 1 | 1 | 0 | 0 | 0 | 1 | 0 | 0 | 0 | S |
| 6 | 0 | 1 | 1 | 1 | 0 | 0 | 0 | 1 | 0 | 1 | 1 | $G_2$ |
| 7 | 0 | 0 | 0 | 1 | 0 | 0 | 1 | 1 | 0 | 1 | 1 | M |
| 8 | 0 | 0 | 0 | 0 | 0 | 1 | 1 | 0 | 0 | 1 | 1 | M |
| 9 | 0 | 0 | 0 | 0 | 0 | 1 | 1 | 0 | 1 | 1 | 1 | M |
| 10 | 0 | 0 | 0 | 0 | 0 | 1 | 1 | 0 | 1 | 0 | 1 | M |
| 11 | 0 | 0 | 0 | 0 | 1 | 1 | 1 | 0 | 1 | 0 | 0 | M |
| 12 | 0 | 0 | 0 | 0 | 1 | 1 | 0 | 0 | 1 | 0 | 0 | $G_1$ |
| 13 | 0 | 0 | 0 | 0 | 1 | 0 | 0 | 0 | 1 | 0 | 0 | Stationary $G_1$ |

**Table 2.** Temporal evolution of protein states for the simplified cell-cycle network of Fig. 1B. The right column indicates the cell-cycle phases. Note that the number of time steps in each phase do not reflect its actual duration.

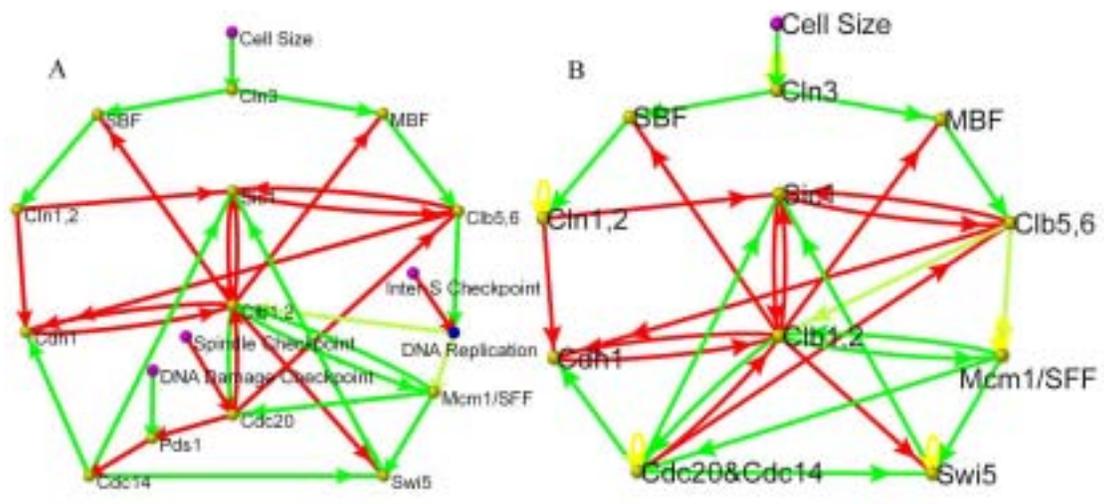

**Fig. 1.** (**A**) The cell-cycle network of the budding yeast. (**B**) Simplified cell-cycle network with only one checkpoint "cell size".

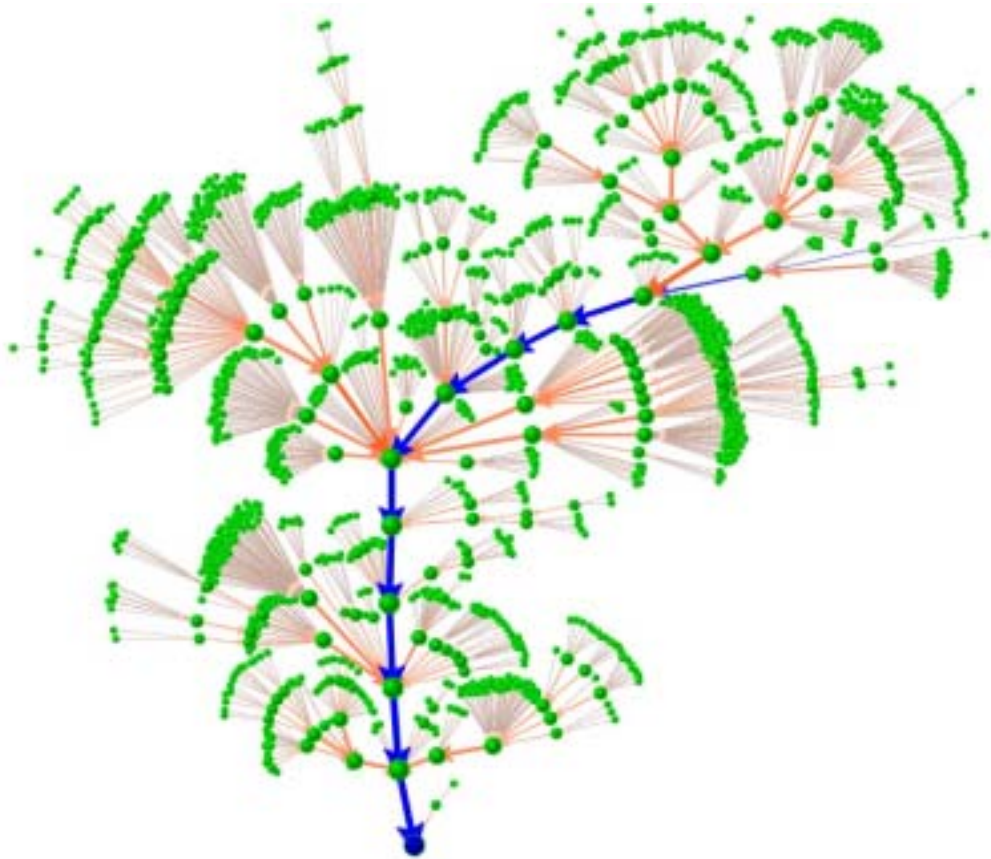

**Fig. 2.** Dynamical trajectories of the 1764 protein states (green nodes) flowing to the $G_1$ fixed point (blue node). Arrows between states indicate the direction of dynamic flow from one state to another. The cell-cycle sequence is colored blue. The size of a node and the thickness of an arrow are proportional to the logarithm of the traffic flow passing through them.

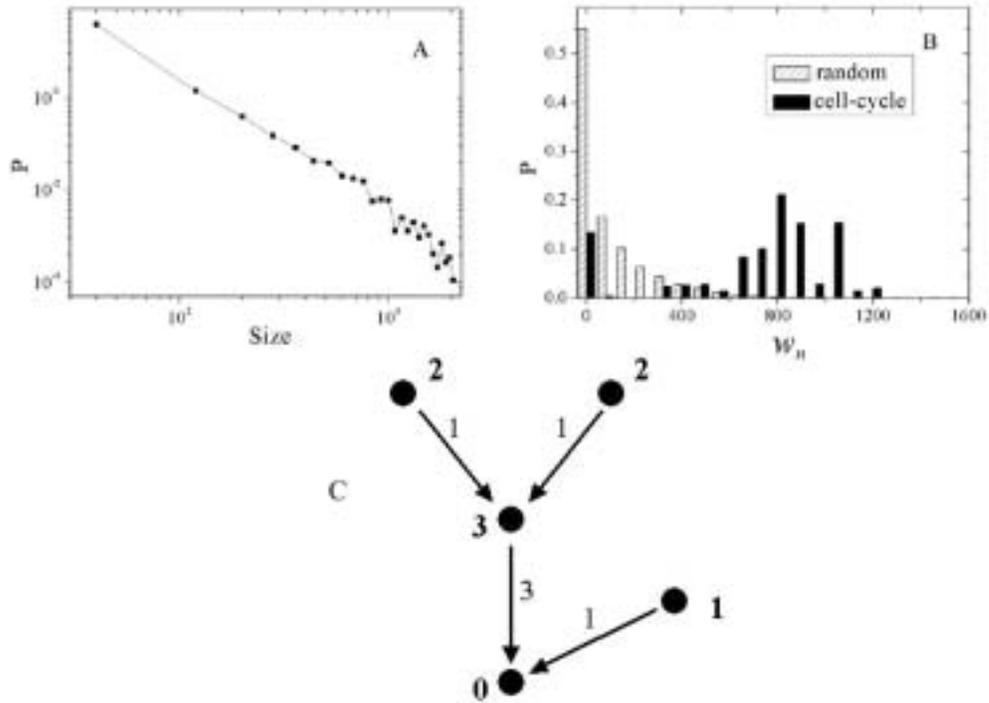

**Fig. 3.** Comparison with random networks. (**A**) Attractor size distribution of random networks. (**B**) $w$-value distributions for the cell-cycle network and for random networks. 10000 random networks were used to generate the statistics. (**C**) Schematic illustration of the definition of $w_n$. The number next to an arrow indicates the total traffic through the arrow. The number next to a node is the $w_n$ of the node.

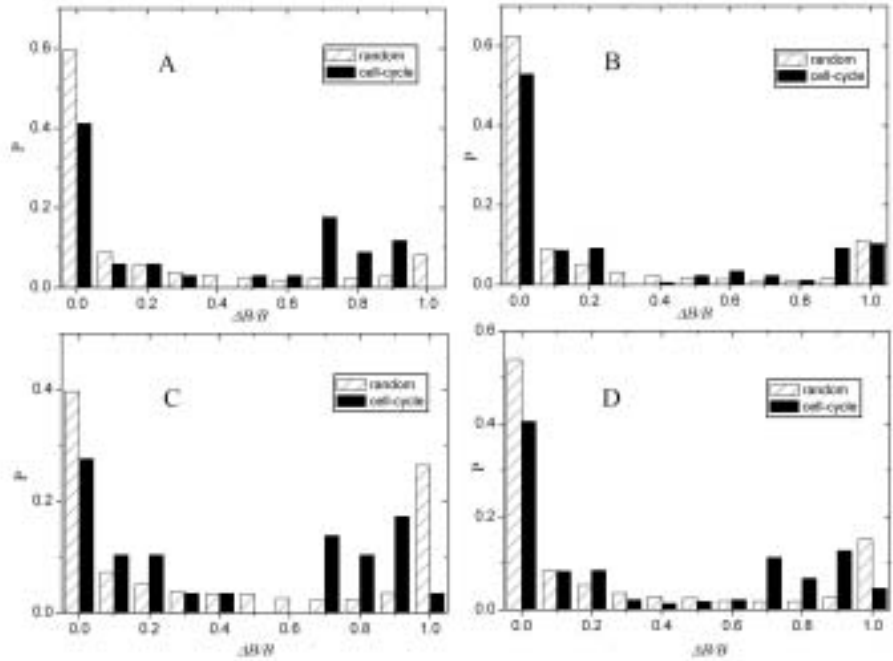

**Fig. 4.** The histogram of the relative changes of the size of the basin of attraction for the biggest fixed point with respect to network perturbations. (**A**) 34 line deletions; (**B**) 174 line additions; (**C**) 29 red-green switchings; and (**D**) the average of **A**-**C**. 1000 random networks were used to generate statistics.

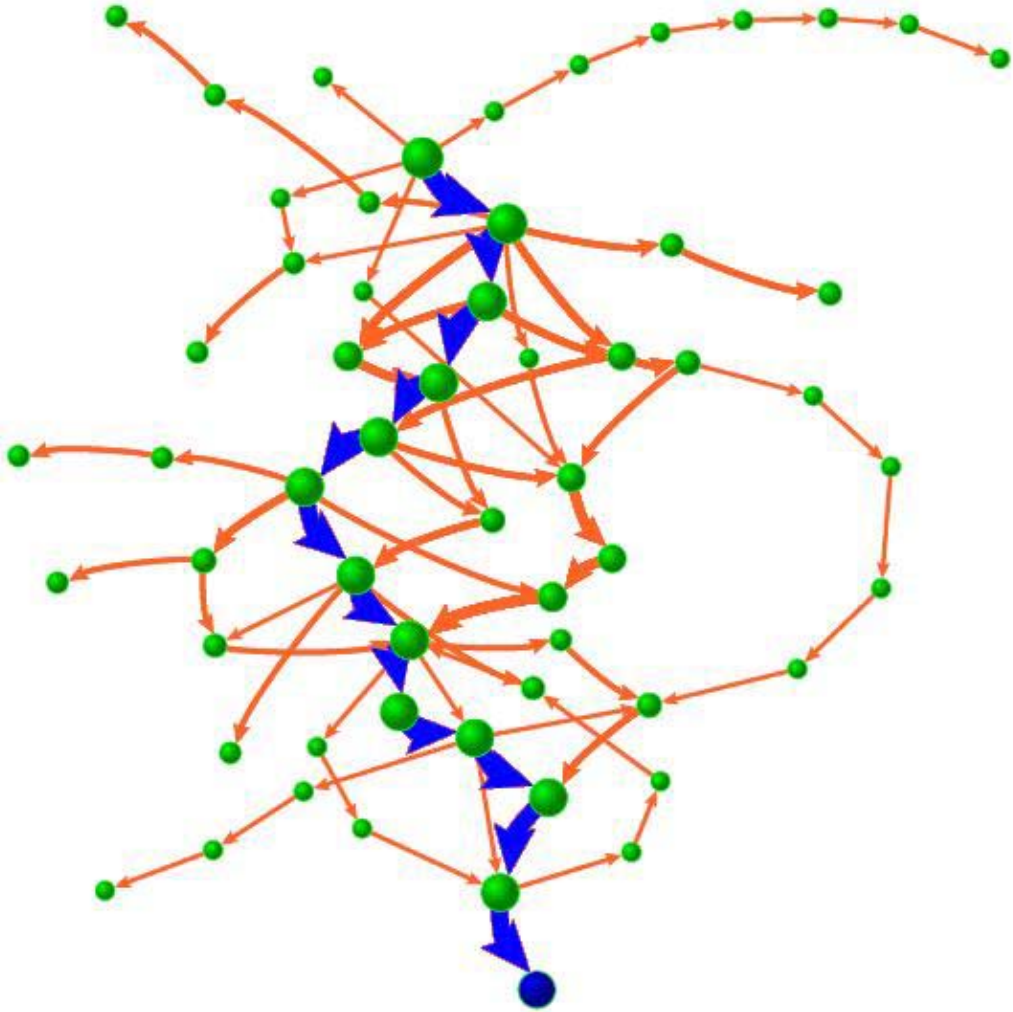

**Fig. 5.** Trajectories of the perturbed cell-cycle network starting from the START. The trajectories from each kind of perturbations (34 from arrow deletions, 174 from arrow additions, and 29 from red-green switchings) are first superimposed on top of each other to form three groups. The three groups are then superimposed on top of each other with equal weights. The width of an arrow and the size of a node are proportional to the logarithm of the number of shared trajectories. The biological pathway is colored blue. The percentages of the perturbed networks that still evolve to the $G_1$ state from START are 41.2%, 57.4%, and 64.7% for arrow-deletion, arrow addition and color-switching, respectively.

**Supporting Information**

**0. The Protein Network in Budding Yeast**

Budding yeast *Saccharomyces cerevisiae* is a widely studied single-celled eukaryotic model organism. Comprehensive protein-protein interaction maps in budding yeast have been rapidly accumulating (S0-1~S0-10), providing an opportunity to study the global or systemic properties of the protein networks.

S0-1.  Uetz, P., Giot, L., Cagney, G., Mansfield, T. A., Judson, R. S., Knight, J. R., Lockshon, D., Narayan, V., Srinivasan, M., Pochart, P., *et al.* (2000) *Nature* **403**, 623−627.
S0-2.  Schwikowski, B., Uetz, P. & Fields, S. (2000) *Nat. Biotechnol.* **18**, 1257-1261.
S0-3.  Ito, T., Chiba, T., Ozawa, R., Yoshida, M., Hattori, M. & Sakaki, Y. (2001) *Proc. Natl. Acad. Sci. USA* **98**, 4569-4574.
S0-4.  Ideker, T., Thorsson, V., Ranish, J. A., Christmas, R., Buhler, J., Eng, J. K., Bumgarner, R., Goodlett, D. R., Aebersold, R. & Hood, L. (2001) *Science* **292**, 929-934.
S0-5.  Tong, A. H., Evangelista, M., Parsons, A. B., Xu, H., Bader, G. D., Page, N., Robinson, M., Raghibizadeh, S., Hogue, C. W., Bussey, H., *et al.* (2001) *Science* **294**, 2364-2368.
S0-6.  Gavin, A. C., Bosche, M., Krause, R., Grandi, P., Marzioch, M., Bauer, A., Schultz, J., Rick, J. M., Michon, A. M,, Cruciat, C. M., *et al.* (2002) *Nature* **415**, 141-147.
S0-7.  Ho, Y., Gruhler, A., Heilbut, A., Bader, G. D., Moore, L., Adams, S. L., Millar, A., Taylor, P., Bennett, K., Boutilier, K., *et al.* (2002) *Nature* **415**, 180−183.
S0-8.  Lee, T. I., Rinaldi, N. J., Robert, F., Odom, D. T., Bar-Joseph, Z., Gerber, G. K., Hannett, N. M., Harbison, C. T., Thompson, C. M., Simon, I., *et al.* (2002) *Science* **298**, 799-804.
S0-9.  von Mering, C., Krause, R., Snel, B., Cornell, M., Oliver, S. G., Fields, S. & Bork, P. (2002) *Nature* **417**, 399-403.
S0-10. Tucker, C. L., Gera, J. F. & Uetz, P. (2001) *Trend in Cell Biology* **11**,102-106.
S0-11. There are also many online databases containing rich information about protein functions and interactions in yeast. For example, Munich Information Centre for Protein Sequences (http://mips.gsf.de/), the Yeast Proteome Database (http://www.proteome.com), and the Saccharomyces Genome Database (http://www.yeastgenome.org/).

**1. The network governing the cell-cycle process in budding yeast**

In eukaryotes the cell-cycle process is divided into two phases: interphase and mitosis. Interphase consists of $G_1$, S and $G_2$ phases. Mitosis can be sub-divided into prophase, metaphase, anaphase, and telophase. Chromosomes condense during prophase, align during metaphase, separate during anaphase, and de-condense during telophase. A highly regulated and complex network governs the cell-cycle process. The machinery of cell-cycle control is known in more detail for budding yeast (S1-1) than for any other eukaryotic organisms.

In yeast, cell-cycle initiation is coupled to cell size probably through the accumulation of Cln3 (S1-1, S1-2). The nuclear concentration of the $G_1$ cyclin Cln3 increases with the total cell mass (S1-1, S1-3). When the level of Cln3/Cdc28 complex is larger than a certain threshold, it triggers $G_1$/S transcription by activating (phosphorylating) SBF (Swi4 and Swi6) (S1-4) and MBF (Mbp1 and Swi6) (S1-5). In fact, Bck2 acts genetically as a parallel system to Cln3, activating SBF and MBF. Since the function of Bck2 is not indispensable (S1-2), we use Cln3 to stand for both Cln3 and Bck2 in our network for simplification. SBF and MBF are the transcription factors of CLN1,2 (S1-6) and CLB5,6 (S1-5), respectively. So Cln1,2 and Clb5,6 begin to accumulate. Cln1, Cln2, Clb5 and Clb6 bind to Cdc28 (a CDK protein) respectively to form compounds (S1-7). At first, Clb5,6 accumulate in the inactive trimer form, Clb5/Cdc28/Sic1 and Clb6/Cdc28/Sic1 (S1-8).

While Cln1,2/Cdc28 are active and accumulating, they cause two processes consequentially besides initiating the bud formation: Sic1 is phosphorylated and then degraded by the proteasome—Cln1,2/Cdc28 prime the inhibitor for ubiquitination by SCF+Cdc4 (S1-9); and Cdh1 is inactivated (by phosphorylation) (S1-10, S1-11). Chd1 controls Clb1,2's degradation in M/$G_1$ transition through cooperating with APC (Anaphase Promoting Complex) (S1-12, S1-13). As a result of Sic1's degradation, Clb5,6/Cdc28 become active and drive the cell-cycle process into the S phase (S1-14). Activation of Clb5,6 facilitate the phosphorylation of Sic1 and may have an inhibitive action on Cdc14 (S1-15). If there is no DNA damage, Rad53 will not be activated, otherwise, it will be the effecter of DNA damage checkpoint pathway and arrest the cell cycle in G2 phase (S1-16).

The Mcm1/SFF (Mcm1/Fkh2/Ndd1) complex is the transcription factor of CLB1,2 (S1-17, S1-18, S1-19) and SWI5 (S1-20). The transcription of CLB1,2 also depends on an active mitotic kinase (S1-21). But how Mcm1/SFF begins to work is not exactly known so far. The binding of Ndd1 and the phosphorylation of the complex are key processes. Clb1,2 can phosphorylate the complex (S1-17, S1-18), thus forming a positive feedback loop with Mcm1/SFF. Ndd1, in cooperation with Mcm1/Fkh1,2, is a stage-dependent activator of G2/M phase-specific transcripts (S1-22, S1-23). Presumably, the binding of Ndd1 happens after the successful completion of DNA replication. Here we use a functional node to represent DNA replication in our cell-cycle network, and use two positive arrows (colored light-green to represent uncertainty) from the node to activate the loop. In the simplified cell-cycle network (Fig. 1B), we remove this intermediate node, and use two positive arrows from Clb5,6 to Mcm1/SFF and Clb1,2 to present the indirect action (S1-19). With Sic1 gone and Cdh1 inactivated, the activity of Clb1,2 begins to rise after the S phase. Clb1,2/Cdc28 inactivates SBF (S1-17), so Cln1,2 activity begins to fall. MBF is inactivated by some unknown proteins rather than by Clb1,2 (S1-20), but since MBF and SBF turn off at similar times in the cell cycle under most conditions (S1-25), we assume that MBF is also (indirectly) inactivated by Clb1,2. The inactivation of MBF causes Clb5,6 level to fall. The rising Clb1,2/Cdc28 activity induces mitosis (S1-26).

A key regulator in the metaphase-anaphase transition and in the exit from mitosis is Cdc20 (S1-15). Cdc20 presents substrates to the APC, which is activated by Clb2 (S1-50), for ubiquitination (S1-27). Before metaphase, Cdc20 is inactive, but when DNA is fully replicated and all chromosomes are aligned on the metaphase plate, Cd20 is activated (S1-28, S1-29). The transcription control of CDC20 depends on a similar active mitotic kinase as that involved in the transcription of CLB1,2 (S1-21). There is evidence that Mcm1/SFF plays a role in the transcription of CDC20 (S1-30). Cdc20/APC functions in two aspects: first, it degrades Clb2 (S1-49) and Clb5,6 (S1-12, S1-13); second, it activates Cdh1 and Swi5 by degrading an inhibitor of Cdc14 (S1-13, S1-15, S1-31, S1-32, S1-33). Cdh1, which was inactivated by Cln1,2 through phosphorylation (S1-10, S1-11), conducts Clb1,2 to the APC for ubiquitination (S1-12,S1-13). Swi5 is the transcription factor of SIC1 (S1-34), but in the presence of Clb1,2, it is prevented from entering the nucleus (S1-20). Once Swi5 is in the nucleus, it can increase the amount of Sic1 and thus help to drive the cell to return to the $G_1$ phase. However, Swi5 will not remain for a long time since once it enters the nucleus, it will be destroyed (S1-34). In the simplified network, we combine the three nodes Cdc20, Pds1 and Cdc14 into one node.

Besides the cell-size control, there are several other checkpoint mechanisms in the cell-cycle process, such as the intra-S-phase checkpoint and the DNA-replication and DNA-damage checkpoints to ensure the order of the cell-cycle events and to preserve genomic integrity (S1-35, S1-36, S1-37). The intra-S-phase checkpoint slows DNA replication in response to DNA damage. If replication forks stop in response to intra-S DNA damage, the firing of late origins is blocked by the activity of Mec1 and Rad53 (S1-38, S1-39, S1-40, S1-41). The DNA-damage checkpoint communicates to the mitotic apparatus through Pds1 and Cdc5 by blocking the metaphase to anaphase (M-A) transition (S1-42, S1-43). Pds1 is an anaphase inhibitor, which is required for the DNA damage and spindle checkpoints (S1-44). The spindle assembly checkpoint blocks the metaphase-anaphase transition by inhibiting Cdc20/APC (S1-48).

In the simplified cell-cycle network (Fig. 1B), we add a self-degradation action (yellow line) to the proteins that are not directly negatively regulated by other proteins in the network to simulate their finite life times. This is a simplification for the actual degradation processes. For example, Cln1,2 and Cln3 are degraded by a complex SCF associated with Grr1 (S1-45, S1-46, S1-47), Mcm1/SFF is inactivated when one of its subunit Ndd1 is degraded and the complex is dephosphorylated, and Cdc20 is destructed by APC-dependent proteolysis machinery (S1-21).

| STARTING NODE | ENDING NODE | DESCRIPTION | REFERENCE |
|---|---|---|---|
| Cell Size | Cln3 | The nuclear concentration of Cln3 is proportional to cell mass. When the cell is large enough, Cln3 is "activated". | S1-1, S1-2 |
| Cln3 | SBF | When the level of Cln3/Cdc28 complex is larger than a certain threshold, it triggers G1/S transcription by activating SBF (Swi4 and Swi6). | S1-1, S1-4 |
| Cln3 | MBF | Cln3/Cdc28 complex activates MBF (Mbp1 and Swi6) by the similar mechanism to SBF. | S1-5 |
| SBF | Cln1,2 | SBF is the transcription factor of CLN1,2. | S1-6 |
| MBF | Clb5,6 | MBF is the transcription factor of CLB5,6. | S1-5 |
| Cln1,2 | Sic1 | Cln1,2/Cdc28 complex phosphorylates Sic1 for its degradation. | S1-7 |
| Cln1,2 | Cdh1 | Cln1,2/Cdc28 complex inactivates Cdh1. | S1-13 |
| Sic1 | Clb5,6 | Sic1 binds to Clb5,6/Cdc28 to inactivate the complex's function. | S1-8 |
| Clb5,6 | Sic1 | Clb5,6 phosphorylate Sic1. | S1-15 |
| Clb5,6 | DNA Replication | Clb5,6 initiate DNA replication. | S1-14 |
| Inter-S Checkpoint | DNA Replication | The DNA replication checkpoint in S phase. | S1-36, S1-38~41, |
| Clb5,6 | Cdh1 | Clb5,6 phosphorylate Cdh1 and cause it to be inactive. | S1-15 |
| DNA Replication | Clb1,2 | The transcription of Clb1,2 partly depends on an active mitotic kinase which is related to completion of correct DNA replication. | S1-21 |
| DNA Replication | Mcm1/SFF | The binding of Ndd1, which happens in G2/M, is a key process to activate the complex. | S1-19, S1-23 |
| Clb1,2 | Mcm1/SFF | Clb1,2 phosphorylate the complex. | S1-19 |
| Clb1,2 | Sic1 | Clb1,2 phosphorylate Sic1. | S1-1 |
| Clb1,2 | SBF | Clb1,2/Cdc28 inactivate SBF. | S1-17 |
| Clb1,2 | MBF | MBF is inactivated by some unknown proteins rather than Clb1,2, but since MBF and SBF turn off at similar time in cell cycle under most conditions, we add a red line from Clb1,2 to MBF to represent this logic relation. | S1-17, S1-25 |
| Sic1 | Clb1,2 | Clb1,2/Cdc28 are inactivated by the binding of Sic1. | S1-8 |
| Mcm1/SFF | Clb1,2 | Mcm1/SFF is the transcription factor of CLB1,2. | S1-17, S1-18 |
| Mcm1/SFF | Cdc20 | The transcription control of CDC20 depends on Mcm1/SFF to some extent. | S1-30 |
| Cdc20 | Clb5,6 | Cdc20 presents Clb5,6 to the APC for ubiquitination. | S1-12, S1-13 |
| Clb1,2 | Swi5 | Clb1,2 phosphorylate Swi5 to prevent its entry to the nucleus. | S1-20 |
| Cdc20 | Pds1 | Cdc20 targets Pds1 for destruction to promote separation of sister chromatids. | S1-15 |
| DNA Damage Checkpoint | Pds1 | Pds1 is phosphorylated in response to DNA damage. | S1-42, S1-43 |
| Pds1 | Cdc14 | When Pds1 is present, Cdc14 is bound to nucleolar RENT complex. | S1-15 |
| Cdc14 | Sic1 | Cdc14 can dephosphorylate and activate Sic1. | S1-15 |
| Cdc14 | Cdh1 | Cdc14 can dephosphorylate and activate Cdh1. | S1-15 |
| Cdc14 | Swi5 | Cdc14 dephosphorylates Swi5. | S1-33 |
| Cdh1 | Clb1,2 | Cdh1 controls degradation of Clb1,2. | S1-12 |
| Swi5 | Sic1 | Swi5 is the transcription factor of SIC1. | S1-34 |
| Mcm1/SFF | Swi5 | Mcm1/SFF is the transcription factor of SWI5. | S1-20 |
| Clb1,2 | Cdc20 | Clb2 activates APC via phosphorylation | S1-50 |
| Cdc20 | Clb1,2 | Cdc20/APC degrades Clb1,2 | S1-49 |
| Spindle checkpoint | Cdc20 | The spindle assembly checkpoint inhibits Cdc20/APC | S1-48 |
| Clb1,2 | Cdh1 | Clb1,2 inactivate Cdh1 by phosphorylation | S1-10 |

**Table S-1.** References to each interaction in Fig. 1A.

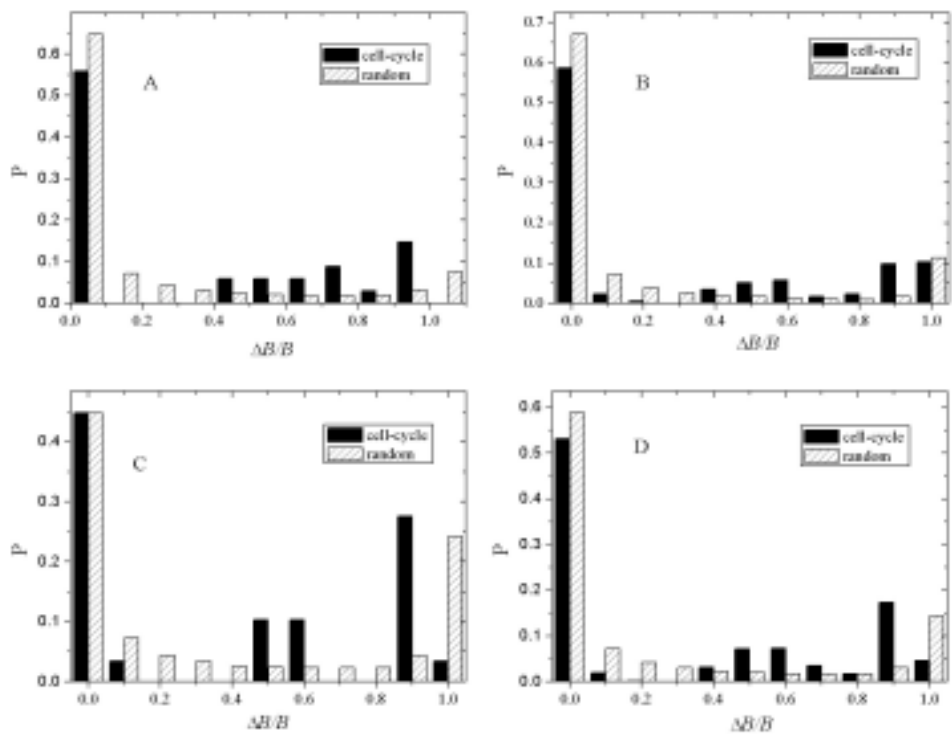

**Fig. S-1.** The histogram of the relative changes of the size of the basins of attraction for the biggest fixed point with respect to network perturbations, with $a_g=1$, $a_r=-10$, and $t_d=4$. (**A**) 34 line deletions; (**B**) 174 line additions; (**C**) 29 red-green switchings; and (D) the average of **A**-**C**. 1000 random networks were used to generate statistics.

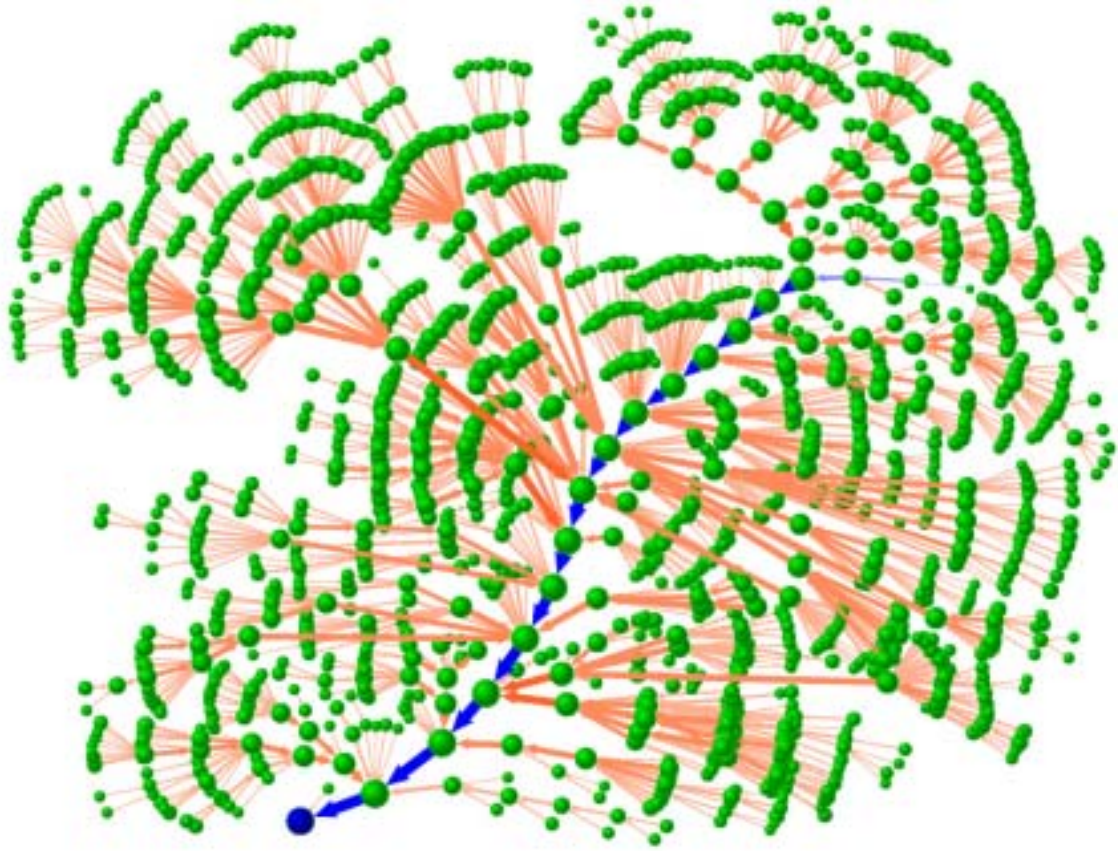

**Fig. S-2.** The trajectories to the $G_1$ fixed point for the full cell-cycle network (Fig. 1), but with only one checkpoint: cell size, and with $a_g=1$, $a_r=-1$, and $t_d=1$. Out of all $2^{14}=16384$ initial states, 14884 of them (90.8%) flow to the $G_1$ fixed point. For clarity, leaf nodes (nodes without incoming traffic) are omitted from the graph.